\documentstyle[prl,aps,epsf]{revtex}

\begin{document}

\begin{center}
{\Large\bf
Non-factorized genuine twist $3$ in exclusive electro-production
of vector mesons}
\end{center}

\begin{center}
I.V.~Anikin
\\
{\it Bogoliubov Laboratory of Theoretical Physics,
JINR, Russia }
\\
\vspace{0.3cm}
O.V.~Teryaev
\\
{\it Bogoliubov Laboratory of Theoretical Physics,
JINR, Russia}
\end{center}

\begin{abstract}

\noindent
We present an analysis of genuine twist-$3$ quark contributions
to the amplitude of exclusive electro-production of transversely polarized
vector mesons. Using the formalism based on the momentum
representation we calculated all the genuine twist-$3$ terms
of quark contributions to $\gamma_T^*\to\rho_T$
amplitude. We found that these terms can not be
factorized owing to the existence of the infrared divergencies in
the amplitude of hard sub-processes.
\end{abstract}
\vspace{1.cm}

The electro-production of transverse polarized vector mesons
${\rm hadron}(p_1)+\gamma^*(q)\to\rho(p)+{\rm hadron}(p_2)$
provides the well-known example
of QCD factorization breaking.
Indeed, the factorization is valid in the
case of longitudinally polarized $\rho$-meson production unless the
high twist effects are included \cite{Col}.
However,
the description of transversely polarized meson production
is strongly complicated due to the existence of
infrared divergencies in amplitudes, breaking down
the factorization (see e.g. \cite{Man2}, \cite{Rad}
and references therein).

The amplitude
of transverse vector meson production corresponds to the $1/Q$ suppressed
contributions in comparison with the longitudinal vector meson case
\cite{Col}.
At the same time, the resent experiments show that the transverse
mesons production amplitudes do sensible contributions even at moderate
virtualities $Q^2$ \cite{Her}.
So, to describe these processes we are forced to deal with
the taking into account the terms of $1/Q$ -- power while,
on the other hand,
it should yield to the problems related to the factorization theorem.
In Ref. \cite{Man1}, the analysis of twist-$2$
amplitudes for hard exclusive lepto-production of mesons in
terms of generalized parton distributions (GPDs or SPDs) was presented.
Recently authors of \cite{Man2} discussed the factorization problems
in the electro-production of light vector mesons from transversely
polarized photons. They have taken into account just kinematical
twist-$3$ terms.
Also the helicity flip amplitude of transversely polarized vector
mesons production was considered within the kinematical
Wandzura--Wilczek
approximation \cite{Kiv}.

Although the kinematical and dynamical (genuine) higher twists
contributions are, generally speaking, independent, there
are notable exceptions in  some kinematical regions.
In Deep Inelastic scattering at $x_B\to 1$ the
kinematical higher twist terms, described by Nachtmann
variable, lead to inconsistencies unless the genuine
higher twists are taken into account \cite{DGP}.
One cannot exclude, that the genuine higher twists
may cure, at least partially, the problem arising
from the treatment of the end-point regions in
hard electro-production.

Thus we want to study the role of genuine twist-$3$ contributions in the
factorization theorem breaking.
We adhere the approach based on
the momentum representation the basic stages of which are expounded in
previous papers \cite{Ani1}-\cite{Ani3}.

We now derive the electro-production amplitude of
transversely polarized $\rho$-meson in terms of the coefficient
function and "soft" functions.
First let us introduce the kinematics in such a manner:
the $p$ is momentum of transversely polarized $\rho$-meson and
its polarization vector is $e^T$; the momentum of virtual photon is denoted
by $q (Q^2=-q^2)$.
We assume that the initial hadron momentum $p_1$
and final hadron momentum $p_2$ are collinear, and
$p_1^2=p_2^2=t=0$,
neglecting all the relevant higher twists contributions.
In addition we neglect squares of meson
masses, restricting ourselves to twist 3 contributions.
With the help of $p_1$ and $p_2$ momenta we build the general relative momentum
$\overline{P}=(p_2+p_1)/2$ and transfer momentum $\Delta=p_2-p_1$.
In our approximation
$\overline{P}^2= \Delta^2=0$.

We introduce the parameterizations of all
the matrix elements needed for calculations of amplitudes.
We use the axial gauge $n\cdot A=0$, where $n^2=0, n\cdot p=1$
In terms of the light-cone basis vectors
the $\rho$-meson--to--vacuum matrix elements can be written as
(keeping just the terms up to the twist-$3$ order):
\begin{eqnarray}
\label{par1}
&&\langle p|\bar\psi(0)\gamma_{\mu} \psi(z)|0\rangle
\stackrel{{\cal F}}{=}
\varphi_1(y)p_{\mu}+\varphi_3(y)e^{T}_{\mu},
\langle p|
\bar\psi(0)\gamma_{\mu}
\stackrel{\longleftrightarrow}
{\partial^T_{\rho}} \psi(z)|0 \rangle
\stackrel{{\cal F}}{=}
\varphi_1^T(y)p_{\mu} e^T_{\rho},
\\
\label{par1.1}
&&\langle p|
\bar\psi(0)\gamma_5\gamma_{\mu} \psi(z) |0\rangle
\stackrel{{\cal F}}{=}
i\varphi_A(y)\varepsilon_{\rho\alpha\beta\delta}
e^{T}_{\alpha}p_{\beta}n_{\delta},
\nonumber\\
&&\langle p| \bar\psi(0)\gamma_5\gamma_{\mu}
\stackrel{\longleftrightarrow}
{\partial^T_{\rho}} \psi(z) |0\rangle
\stackrel{{\cal F}}{=}
i\varphi_A^T (y) p_{\mu}\varepsilon_{\rho\alpha\beta\delta}
e^{T}_{\alpha}p_{\beta}n_{\delta},
\\
\label{par1.2}
&&\langle p|
\bar\psi(0)\gamma_{\mu}g A_{\rho}^T(z_2) \psi(z_1) |0\rangle
\stackrel{{\cal F}}{=}
\Phi(y_1,y_2)p_{\mu} e^{T}_{\rho},
\nonumber\\
&&\langle p|
\bar\psi(0)\gamma_5\gamma_{\mu} g A_{\rho}^T(z_2) \psi(z_1) |0\rangle
\stackrel{{\cal F}}{=}
i J(y_1,y_2)p_{\mu}\varepsilon_{\rho\alpha\beta\delta}
e^{T}_{\alpha}p_{\beta}n_{\delta},
\end{eqnarray}
where $\stackrel{{\cal F}}{=}$ denotes the Fourier transformation
with measure ($z_i=\lambda_i n$)
\begin{eqnarray}
dy \,e^{ -iy\,pz}
&&\quad {\rm for\,\, quark\,\, correlators},
\nonumber\\
dy_1 \,dy_2 \,e^{ -iy_1\,pz_1
-i(1-y_2 - y_1)\,pz_2 }
&&\quad {\rm for\,\, quark-gluon\,\, correlators}.
\end{eqnarray}

In the adopted collinear limit we directly obtain that
$p_2=(1-\xi)/(1+\xi)p_1=\kappa p_1$.
Therefore,
$u(p_2)=\sqrt{\kappa} u(p_1)$. This lead to the symmetrical
nucleon spinor forms like $ \bar u(p_1)\sqrt{\kappa}\hat n u(p_1)$
whose calculation is straightforward.
As a result we are able to introduce the parameterization
of relevant matrix elements for nucleons.
Further, keeping the twist-$2$ terms only, we write
\begin{eqnarray}
\label{parhad-q}
\langle p_2|\bar\psi(0)\gamma_{\mu} \psi(\tilde z)
 | p_1\rangle
\stackrel{{\cal F}}{=}
H_1(x)\overline{P}_{\mu}
\end{eqnarray}
\noindent
for the nucleon--nucleon matrix element of
pure quark correlator .

In this paragraph we compute
the quark contributions to the production amplitude
which are generated by the diagrams with the quark legs
coming from the nucleon blob.
The calculating of the simplest Feynman diagrams
give the following expression:
\begin{eqnarray}
\label{qdsim}
{\cal A}_{1,\,\mu}^{(q),\,\gamma_T^*\to\rho_T}&=&
4\,\frac{C_F}{N_c}\,\frac{e_{\mu}^T}{Q^2}
\int_{-1}^{1}dx H_1(x)\biggl[
\frac{1}{x-\xi+i\epsilon}-\frac{1}{x+\xi-i\epsilon}
\biggr]
\nonumber\\
&&\int_{0}^{1}\frac{dy}{y(1-y)}
\Biggl( \varphi_A(y)+\varphi_3(y)\Biggr).
\end{eqnarray}
Next, summing the diagrams that do contribute to the genuine
twist-3, we finally obtain the
expression for the quark distribution amplitude. We write
\begin{eqnarray}
\label{qdtw3}
{\cal A}_{2,\,\mu}^{(q),\,\gamma_T^*\to\rho_T}&=&
4\,\frac{C_F}{N_c^2-1}\,\frac{e_{\mu}^T}{Q^2}
\Biggl\{
\xi\int_{-1}^{1}dx H_1(x)\biggl[
\frac{1}{(x+\xi-i\epsilon)^2}+\frac{1}{(x-\xi+i\epsilon)^2}
\biggr]
{\cal I}_1^{(q)} +
\Biggl.
\nonumber\\
\Biggr.
&& \int_{-1}^{1}dx H_1(x)\biggl[
\frac{1}{x-\xi+i\epsilon}-\frac{1}{x+\xi-i\epsilon}
\biggr]
{\cal I}_2^{(q)}
\,\Biggr\},
\end{eqnarray}
where
\begin{eqnarray}
\label{I1q}
&&{\cal I}_1^{(q)}=\int_{0}^{1} dy_1\,dy_2
\Biggl\{
\tilde J(y_1,y_2)\biggr[
4C_F\biggl( \frac{1}{(1-y_1)^2}-\frac{1}{(1-y_2)^2}\biggr)+
\biggr.
\Biggr.
\nonumber\\
\Biggl.
\biggl.
&&C_A\biggl( \frac{1}{y_2(1-y_1)}-\frac{1}{y_1(1-y_2)}\biggr)+
\frac{2C_F-C_A}{y_1+y_2}
\biggl( \frac{1}{y_1}-\frac{1}{y_2}\biggr) \biggr]-
\Biggr.
\nonumber\\
\Biggl.
&&\tilde \Phi(y_1,y_2)\biggr[
4C_F\biggl( \frac{1}{(1-y_1)^2}+\frac{1}{(1-y_2)^2}\biggr)+
C_A\biggl( \frac{1}{y_2(1-y_1)}+\frac{1}{y_1(1-y_2)}\biggr)+
\biggr.
\Biggr.
\nonumber\\
\Biggl.
\biggl.
&&\frac{2C_F-C_A}{y_1+y_2}
\biggl( \frac{1}{y_1}+\frac{1}{y_2}\biggr) \biggr]
\Biggr\};
\end{eqnarray}
\begin{eqnarray}
\label{I2q}
&&{\cal I}_2^{(q)}=\int_{0}^{1} dy_1\,dy_2
\Biggl\{
\tilde J(y_1,y_2)\biggr[
C_F\biggl( \frac{1}{y_2(1-y_1)}-\frac{1}{y_1(1-y_2)}\biggr)+
\frac{2C_F-C_A}{y_1+y_2}
\biggl( \frac{1}{y_2}-\frac{1}{y_1}\biggr) \biggr]-
\Biggr.
\nonumber\\
\Biggl.
&&\tilde \Phi(y_1,y_2)\biggr[
C_F\biggl( \frac{1}{y_2(1-y_1)}+\frac{1}{y_1(1-y_2)}\biggr)+
\frac{2C_F-C_A}{y_1+y_2}
\biggl( \frac{1}{y_2}+\frac{1}{y_1}\biggr) \biggr]
\Biggr\}.
\end{eqnarray}
Here we introduced the notation for
the general parameterizing functions $\tilde\Phi$ and $\tilde J$
consisting of the kinematical and genuine (dynamical) twist-$3$ parts,
{\it i.e.}, for example,
\begin{eqnarray}
\tilde\Phi(y_1,y_2)=\varphi_1^T(y_1)\delta(y_1-y_2)+\Phi(y_1,y_2).
\end{eqnarray}
The structure integrals  of the quark distribution amplitude possess
poles of second order also. This fact lead to the vulnerability
of factorization theorem unless the functions
$\tilde\Phi$ and $\tilde J$ vanish at $y_i\to 1$ or $y_i\to 0$
more rapidly then the first power of $(1-y_i)$ or $y_i$.
Within our formalism we have reproduced, also, the results for
the kinematical twist 3
gluon distribution amplitude obtained in \cite{Man1}.

In summary, we have computed the total quark contributions
to the transversely polarized $\rho$-meson electro-production.
We have found the novel singular terms of
the genuine twist-$3$ to the quark contribution amplitude,
which must be included in the general analysis
of problems related with the factorization theorem breaking.

Also, we would like to thank to A.V. Efremov, N. Kivel, A. Radyushkin,
A. Schaefer, C. Weiss for fruitful discussions and comments.
This work has been supported in part by INTAS (Project 587, call 2000)
and by RFFI Grant 00-02-16696.

\end{document}